\documentclass[times,10pt,twocolumn]{article} 
\usepackage{latex8}
\usepackage{times}
\newcommand{\nfrac}[2]{{#1}/{#2}}
\newcommand{\sent}{{\mathsf{sent}}}
\newcommand{\rcvd}{{\mathsf{rcvd}}}
\newcommand{\lost}{{\mathsf{lost}}}
\newcommand{\lsr}{{\mathsf{lsr}}}
\newcommand{\into}{{\mathsf{into}}}
\newcommand{\opt}{{\mathsf{opt}}}
\newcommand{\cpc}{{\mathsf{cap}}}
\newcommand{\val}{{\mathsf{val}}}

\newcommand{\delay}[2]{{\tau_{#1#2}}}

\newtheorem{theorem}{Theorem}
\newtheorem{corollary}{Corollary}
\newtheorem{lemma}[theorem]{Lemma}

\newtheorem{definitions}{Definitions}

\newcommand{\qed}{\vspace{.1em}\noindent\fbox{\rule{%
0em}{.1em}\rule{.1em}{0em}}\vspace{1em}} 

\newenvironment{proof}{%
\noindent{\bf Proof:}\ }{%
\hfill \qed }

\begin{document}
\title{On-Line End-to-End Congestion Control}
\author{
Naveen Garg\\
Indian Institute of Technology\\
New Delhi, India\\
naveen@cse.iitd.ernet.in\\
  \and
  Neal E.\ Young\\
    Akamai Technologies\\
    Cambridge, MA, USA\\
    neal@young.name\\
}
\maketitle
\thispagestyle{empty}
\begin{abstract}
  
  Congestion control in the current Internet is accomplished
  mainly by TCP/IP.  To understand the macroscopic network
  behavior that results from TCP/IP and similar end-to-end
  protocols, one main analytic technique is to show that the
  the protocol maximizes some global objective function of the
  network traffic.
  
  Here we analyze a particular end-to-end, MIMD
  (multiplicative-increase, multiplicative-decrease) protocol.
  We show that if all users of the network use the protocol,
  and all connections last for at least logarithmically many
  rounds, then the total weighted throughput (value of all
  packets received) is near the maximum possible.  Our
  analysis includes round-trip-times, and (in contrast to most
  previous analyses) gives explicit convergence rates, allows
  connections to start and stop, and allows capacities to
  change.
\end{abstract}

\Section{Congestion control and optimization}

Congestion control in the current Internet is accomplished
mainly by TCP/IP --- 90\% of Internet traffic is TCP-based
\cite{low02internet}.  Meanwhile the design and analysis of
TCP and other end-to-end congestion-control protocols are only
partially understood and are becoming the subject of
increasing attention
\cite{johari00mathematical,karp00optimization}.  One main
analytic technique is to interpret the protocol as solving
some underlying combinatorial optimization problem on the
network --- to show that the protocol causes the traffic
distribution, over time, to optimize some global objective
function
\cite{low02internet,kelly99mathematical,johari01end,athuraliya00optimization,low99optimization,golestani98class,kelly98rate,shenker90theoretical,chiu89analysis,paganini01scalable,massoulie00stability,vinnicombe00stability}.

For example, a continuous analogue of TCP-Reno (under various
assumptions about the network) maximizes \(\sum_i \tau_i^{-1}
\arctan(\tau_i x_i)\), where $\tau_i$ is the (constant)
round-trip time of packets sent by the $i$th user and the
variable $x_i$ is that user's transmission rate
\cite{low02internet}.  (Each term in the sum is a smoothed
threshold function.)  Similarly, a continuous analogue of
TCP-Vegas maximizes \(\sum_i \alpha_i d_i \log x_i\) where
$d_i$ is the round-trip propagation delay of packets sent by
the $i$th user and $\alpha_i$ is a protocol parameter.
Typically these results concern a continuous analogue of the
protocol.  They analyze a system of differential equations
where time is continuous
and each rate $x_i$ is a continuous function of time.  They
show (e.g.\ using a Lyapunov function
\cite{hochstadt75differential,sanchez68ordinary}) that, as
time tends to infinity, the vector $x$ tends to an equilibrium
point that maximizes the objective function in question.

This approach is very general.  It has been used to design and
analyze protocols other than TCP, including protocols that
require the network routers to explicitly transmit congestion
information to the users by means other than packet loss and
latency.  (Typically the congestion signals are dual variables
--- Lagrange multipliers, or ``shadow prices'' --- with
interesting economic interpretations.)  For a survey of
results of this kind, see
\cite{low02internet,kelly99mathematical}.  A few specific
technical papers include
\cite{johari01end,athuraliya00optimization,low99optimization,golestani98class,kelly98rate,shenker90theoretical,chiu89analysis}.

\begin{figure*}[t]
  \begin{center}
\fbox{\parbox{0.97\textwidth}{

  \medskip

  {\bf Inputs for connection $p$:}
  starting rate $f_0(p)>0$, active time interval $T_p = \{s_p,s_p+1,\ldots,e_p\}$.

  {\bf Parameters: $\alpha_p > 0$, $\beta_p \in (0,1)$}

  \medskip
    
  At times $t = s_p,s_p+1,\ldots,s_p+\delay p{}$ take $\sent(p,t) := f_0(p)$.

  At times $t=s_p+1+\delay p{},\ldots,e_p$, take
  \[
  \sent(p,t)
  \,:=\, \sent(p,t-1-\delay p{}) \times\big[{1 + \alpha_p   -   \beta_p \lsr(p,t-1)}\big].
  \]
    }}
    
    \caption{The Linear MIMD Protocol.
      In round $t$ on path $p$,
      $\lsr(p,t)$ is the observed packet loss fraction,
      $\delay p{}$ is the round-trip time.
      For the main result, $\beta_p \in (0,\epsilon]$
      and $\alpha_p = \beta_p\epsilon\,\val(p)$.
      }
    \label{fig:mimd}
  \end{center}
\end{figure*}

Our interest in this paper is in protocols that are both
on-line and {\em end-to-end}: the number of packets sent on a
path $p$ at time $t$ is determined solely by the number of packets
sent and received on $p$ in previous rounds.  (The protocol has
no a-priori knowledge of the network or how the paths relate
to it, and learns about the network only through packet loss.)
Such protocols are implementable in the current Internet,
without modifications to routers.
The protocol we analyze in this paper, which we call the
Linear MIMD Protocol, is an example (see Fig.~\ref{fig:mimd}).
It can be implemented by modifying only the TCP server.

Generally, existing works (that formally analyze end-to-end
protocols implementable in the current Internet) assume that
all connections start at time 0 and continue indefinitely in a
static network.  They show that the objective function is
optimized in the limit as time tends to infinity.  (See Low's
survey and Kelly's survey \cite{kelly99mathematical}.)  The
only exceptions that we are aware of are for the special case
of a single-bottleneck network
\cite{gorinsky00additive,chiu89analysis}.  Thus, we do not yet
have a complete theoretical understanding of speed of convergence
(noted as important by Low in his survey \cite{low02internet})
or of the effects of dynamic connections and changing network
conditions.  (These issues have been studied 
empirically, e.g.\ \cite{bansal01dynamic}.)

Here we use a relatively dynamic model: time is discrete,
connections start and stop, network capacities vary with
time.  The objective function we study is {\em total weighted
  throughput} (the total value of all packets delivered).

Our main result
is that, for some small $T$, {\em as long as each connection
  lasts at least $T$ rounds}, the protocol in
Fig.~\ref{fig:mimd} achieves a total weighted throughput of at
least $(1-\epsilon)\opt$.  Here $\opt$ is the maximum possible
weighted throughput of any solution that respects capacity
constraints and assigns a {\em fixed} rate $f(p)$ to each path
$p$ while the path is active. $T$ is proportional to a
logarithmic term over $\epsilon^3$.  To sidestep the question
of how the protocol finds a reasonable {\em starting} rate for
each path, we analyze the speed of convergence given arbitrary
(feasible) initial rates.

Most existing works assume instantaneous feedback about
congestion.  In practice feedback is delayed due to round-trip
times, but delayed feedback is harder to analyze formally.
Some recent works such as
\cite{paganini01scalable,massoulie00stability,vinnicombe00stability}
study the effect of delayed feedback on convergence, and even
suggest that some variants of TCP may become unstable as
network capacity becomes large \cite{low02dynamics}.  Here we
do model delayed feedback, although we assume the delay on
each path $p$ is a {\em fixed} constant $\delay p{}$.  We show
that the convergence rate of the protocol grows linearly with
delay.

Existing works model packet loss in the network in various
ways.  (Some also model variable latency due to queuing,
which we do not.)  Roughly, we assume that a network resource
(switch, router, etc.) discards packets only if congested, and
then in an approximately {\em fair} manner --- so that no path
incurs significantly disproportional loss {\em over time} (see
Condition~(\ref{eq:fairLoss}) later in the paper).  We believe
the packet-discard model is realistic in practice.  (See the
final section.)

Here is a formal statement of our result.  Let $\delay p{}$ be
the round-trip delay on path $p$.  Note that $\alpha_p$ and
$\beta_p$ are parameters of the protocol and $f_0(p)$ is the
initial sending rate on path $p$.

\newcommand{\mainresult}{
  Assume each $\val(p)\le 1$.
  Fix $\epsilon>0$.
  Assume $\epsilon$-fair loss on each path.
  Let $U_p$ be the maximum amount received in any round on path $p$.
  
  The weighted throughput achieved by the Linear MIMD Protocol
  with $\beta_p = O(\epsilon)$ and
  $\alpha_p=\epsilon\,\beta_p\,\val(p)$ is $(1-O(\epsilon))\opt$
  provided the duration $|T_p|$ of each connection is at least
  \[\Omega\left(\max_p\frac{(1+\delay p{})\ln(U_p/f_0(p))}{\epsilon^2\beta_p\val(p)}\right).\]
}
\smallskip
\noindent{\bf Theorem \ref{theorem:main}}{\em
  \mainresult}%
In today's Internet, a generous upper bound for the ratio
$U_p/f_0(p)$ is around $10^4$ --- the ratio between 100 Mbps
(the most a typical server can transmit) and 10Kbps (about one
packet per second).

Protocols that maximize weighted throughput within a
$1-\epsilon$ factor are necessarily {\em unfair} (they may
allocate little or no bandwidth to some connections).  Also, the
protocol that we study is not innately {\em
  tcp-friendly}, although, with a proper setting of the
parameters, the loss rates the protocol induces in the network
can be made close to the ``background'' level of loss in the
Internet (typically 2-4\% in well-capacitated networks).  For
a discussion of other limitations, and future directions, see
the final section.

The protocol can be tuned to network conditions.  For example,
in a network with low packet-loss rates, a particularly simple
special case of the protocol --- ``for each packet received,
send $1+\epsilon$ packets'' --- converges faster by a factor
of $1/\epsilon$.  The analysis also generalizes to the case
when each path $p$ can use a resource $r$ to some extent
$M_{pr} \in [0,1]$.  (These results are omitted from the
proceedings version.)

\paragraph*{Application: multi-path bandwidth testing.}

Suppose a web site has multiple servers hosted in an Internet
data center.  The web site owner has a contract with the
network provider for 100 Mbps access to the Internet.  The
servers send traffic through multiple, possibly shared,
up-links, and then 
through the provider's local network, toward a backbone
and/or toward the provider's local users.  How can the web
site owner find an optimal traffic distribution, one that
maximizes the traffic sent from its servers while respecting
the local bandwidth constraints?  How can the site owner
verify that the maximum possible traffic is at least 100 Mbps
(or prove to the network provider that 100 Mbps is not
possible, in violation of the contract)?  This is the problem
that motivated this work.  It involves multiple paths, in
contrast to the more-well studied problem of estimating
bandwidth along a single path
\cite{lai99measuring,keshav93controltheoretic,lai01nettimer,lai00measuring,carter96measuring}.

The Linear MIMD Protocol (or any protocol that maximizes
global aggregate throughput) can be used to solve this
problem as follows. Select a large representative sample of
paths (from the servers to destination IP's that the servers
serve).  Run the Linear MIMD Protocol simultaneously on the
paths (giving each path equal value) until near-convergence.
The resulting aggregate throughput will be a good estimate of
the maximum multi-path bandwidth.

In practice, it may suffice to send packets only over
appropriate {\em prefixes} of the paths; this requires
packet-programming techniques that we don't describe here.
In our experience, the time it takes for a typical test is on
the order of a minute.  An advantage of this approach is that
it works even in the presence of ``hidden'' bottlenecks, such
as switches, whose presence can be known to the network user
only through packet loss.

Note that a protocol such as TCP generally does not maximize
aggregate throughput, and so doesn't work for this
application.

\paragraph*{Other related work.}

The protocol can be viewed as a new Lagrangian-relaxation
algorithm (adapted to and implemented in the network setting)
for the underlying packing/covering problem.  It is most
similar in spirit to recent works such as
\cite{garg98faster,young01sequential}, which are in turn part
of a large body of work over the last decade
\cite{fleischer00approximating,grigoriadis96approximate,grigoriadis96coordination,karger95adding,leighton95fast,klein94faster,luby93parallel,plotkin95fast,shahroki90maximum}.
The focus of all of these works are on Lagrangian-relaxation
algorithms with provable convergence rates (i.e., running-time
bounds).  Those works, and this one, are technically related
to algorithms for ``boosting'' and ``following expert advice''
in learning theory (e.g.\ \cite{freund99adaptive}).

Within theoretical computer science, works focusing on
distributed optimization of related problems include
\cite{bartal97global,papadimitriou93linear,awerbuch98converging,afek96convergence,korilis92why}.
See also theoretical works on routing and related problems
\cite{awerbuch95routing,awerbuch95control,awerbuch94throughput}.
Works studying related issues of resource allocation in
networks in a game-theoretical spirit include
\cite{jain01applications,roughgarden00how,feigenbaum00sharing,goel00combining,kleinberg99fairness}.

For a critique of the pricing-based research agenda for
congestion control (which is closely related to the
optimization paradigm), see \cite{shenker95pricing}.

\Section{On-line end-to-end network model}

We model a network as a set of {\em resources} (e.g., cables,
routers, switches).  Each resource $r$ has a capacity
$\cpc(r,t)\ge 0$ that can vary over time.  A {\em connection}
in the network is identified with a path $p$, and has a value
$\val(p)\in [0,1]$ and an {\em active time interval} $T_p$ (a
finite contiguous subsequence of the times
$\{0,1,2,\ldots\}$).  The path $p$ represents the fixed route
(determined by the routers) in the network from the source to
the destination.  We identify each path with the ordered
sequence of resources that it uses, and we write $p\sim r$ or
$r \sim p$ if path $p$ uses resource $r$.

To model feedback delay (round-trip times), we assume static
latencies along the paths: $\delay pr \in \{0,1,\ldots\}$
denotes the number of time intervals it takes a packet to
reach its destination after going through resource $r$, while
\(\delay p{}\in\{0,1,\ldots\}\) denotes the number of time
intervals for a packet to travel the entire length of $p$.  
(We assume $\delay p{} \ge \max_{r\sim p} \delay pr$.)
We use the following notations to count the number of the
protocol's packets of various kinds during time $t$:
\begin{center}
\begin{tabular}{rcl}
$\sent(p,t)$ &--& packets injected at path $p$'s source, \\
$\rcvd(p,t)$ &--& packets received at $p$'s destination, \\
$\lost(p,t)$ &--& $\sent(p,t-\delay p{}) - \rcvd(p,t)$,\\
$\into(r,t)$ &--& packets entering resource $r$, \\
$\lost(r,t)$ &--& packets discarded at resource $r$.
\end{tabular}
\end{center}
We assume for simplicity that all packets are the same size,
and (adopting the standard ``fluid'' model) we allow the
number of packets sent, received, lost, etc.\ to take on
arbitrary real values (rather than integer values).  We assume
that the source of each path, by time $t$, knows
$\rcvd(p,t-1)$ (by some mechanism such as TCP
acknowledgments).

We assume that packet loss occurs at a resource $r$ at time
$t$ only if the number of packets entering the resource at
time $t$ exceeds the capacity $\cpc(r,t)$.  If this occurs,
then the loss is adversarial (arbitrary) subject to a {\em
  fair loss} condition, which says that packet loss is not
unduly biased against any particular user.  To explain, note
that if packet loss were perfectly distributed at each
iteration, then for each path $p$ and time $t$ it would be the
case that
\begin{equation}\label{eq:fairLossExact}
  \rcvd(p,t)
  \,=\, 
  \sent(p,t-\delay p{})
  \prod_{r\sim p} \Big[1-\frac{\lost(r,t-\delay pr)}{\into(r,t-\delay pr)}\Big].
\end{equation}
The above would imply
\begin{eqnarray*}
\frac{\lost(p,t)}{\sent(p,t-\delay p{})} 
& = & 1-\frac{\rcvd(p,t)}{\sent(p,t-\delay p{})} \\
& = &  1-\prod_{r\sim p} \Big[1-\frac{\lost(r,t-\delay pr)}{\into(r,t-\delay pr)}\Big]\\
& \le & \sum_{r\sim p} \frac{\lost(r,t-\delay pr)}{\into(r,t-\delay pr)}.
\end{eqnarray*}
The expression on the left-hand side is the fraction of
packets sent on $p$ at time $t-\delay p{}$ that are lost.  One
can interpret the fraction on the right-hand side as the
probability that any particular packet is lost at resource $r$
at the time the packets in question enter it.  Under this
interpretation, the equation above will hold if the fraction
of packets lost on $p$ is at most the expected number.  The
fair loss condition ({\em $\epsilon$-fair loss}) is that this
holds {\em approximately over time}:
\begin{eqnarray}
  \lefteqn{(\forall p)~~
  \sum_{t\in T_p+\delay p{}}
  \frac{\lost(p,t)}{\sent(p,t-\delay p{})}} \nonumber \\ 
   & \le  & (1+\epsilon)
  \sum_{t\in T_p+\delay p{}} \sum_{r\sim p} \frac{\lost(r,t-\delay pr)}{\into(r,t-\delay pr)}. \label{eq:fairLoss}
\end{eqnarray}

We are interested here in protocols that are both {\em
  on-line} and {\em end-to-end}: that is, they determine the
packets sent on a path $p$ at time $t$ as a function of the
values $\{\sent(p,s),\rcvd(p,s):s<t\}$ (and possibly other
path-specific information such as $\val(p)$).  The protocols
have no a-priori knowledge of the network.

The objective is to maximize the total value of the received
packets $\sum_p\sum_{t\in T_p+\delay p{}}\val(p)\rcvd(p,t)$.
The {\em competitive ratio} (a.k.a.\ performance ratio) of the
protocol is the worst-case ratio (over all inputs) of this
quantity to $\opt$ --- the maximum that could by assigning a
fixed sending rate $f(p)$ to each path $p$ while its active
(without exceeding capacity constraints).

\Section{Analysis of the Linear MIMD Protocol}

\begin{definitions}
  Let
  $T'_p = T_p + \delay p{} = \{s_p +\delay p{}, \ldots,e_p+\delay p{}\}$
  and
  lost-to-sent ratio $\lsr(p,t)=\nfrac{\lost(p,t)}{\sent(p,t-\delay p {})}$.
\end{definitions}

\medskip
First, we lower-bound the throughput in terms of the packet loss:
\begin{lemma}\label{lemma:conservative}
  The Linear MIMD Protocol satisfies, for each path $p$,
  \begin{eqnarray*}
  \sum_{t\in T'_p}\rcvd(p,t) 
  & \ge & (\beta_p/\alpha_p -1) \sum_{t\in T'_p} \lost(p,t) \\
  & & \mbox{} - \alpha_p^{-1}(\delay p{}+1)f_0(p).
  \end{eqnarray*}
\end{lemma}
\begin{proof}
  Let $S=\sum_{t\in T'_p} \sent(p,t-\delay p{})$,\\
  $R=\sum_{t\in T'_p}\rcvd(p,t)$,
  and $L=\sum_{t\in T'_p}\lost(p,t)$.
  
  Expanding the product on the right-hand side in the
  definition of the protocol, for $t\in T'_p$,
  \[\sent(p,t+1)
  \le
  (1 + \alpha_p)\sent(p,t-\delay p{}) - \beta_p\lost(p,t).
  \]
  Summing over $t\in T'_p$ gives
  \[\sum_{t\in T'_p} \sent(p,t+1) \,\le \,(1+\alpha_p)\,S- \beta_p L.
  \]
  Subtracting $S$ from both sides and substituting $R+L$ for $S$ gives
  \[
  \alpha_p R \ge
  (\beta_p-\alpha_p) L - (1+\delay p{})f_0(p)
  \]
\end{proof}

Next, we lower-bound the packet loss in terms of $\opt$, but
with a condition.

\begin{definitions}
  Let $\rcvd^*(p)$ denote the number of packets the optimal
  algorithm would send on path $p$ when it is active.  Define
  $\into^*(r,t)$ to be the number of packets the optimal
  algorithm would send through $r$ at time $t$.  Let $\opt =
  \sum_p \val(p)\rcvd^*(p)|T_p|$ denote the optimum weighted
  throughput.
\end{definitions}

\begin{lemma}\label{lemma:dual}
  Assume that only congested resources discard packets.
  If, for some $c$ and each $p$,
  \(\sum_{t\in T'_p} \sum_{r\sim p} \frac{\lost(r,t-\delay
    pr)}{\into(r,t-\delay pr)} \,\ge\,c\,|T_p| \val(p),\)
  then the number of packets lost,
  $\sum_{t} \sum_r \lost(r,t)$, is at least $c\,\opt$.
\end{lemma}
\begin{proof}
  We show that the loss rates at the resources implicitly
  define a solution to the dual linear program of the problem,
  and we use the dual solution to bound $\opt$.
  
  For each of the protocol's packets lost at a resource $r$ at
  time $t$ where $\into^*(r,t)>0$, allocate a charge of
  one credit uniformly across each of the $\into^*(r,t)$
  packets sent through $r$ by $\opt$ at time $t$.  The
  assumption that packets are discarded only by congested
  resources means that $\lost(r,t) > 0$ only if
  $\into(r,t)\ge\cpc(r,t)\ge\into^*(r,t)$, so the charge to
  each packet that $\opt$ sends through $r$ at time $t$ is
  \[
  \frac{\lost(r,t)}{\into^*(r,t)} \,\ge\, \frac{\lost(r,t)}{\into(r,t)}.
  \]
  The number of lost packets is at least the total charge to
  $\opt$'s packets, which by the above (and the supposition in
  the lemma about each path $p$) is at least
  \begin{eqnarray*}
  \lefteqn{\sum_p \rcvd^*(p) \sum_{r\sim p} \sum_{t\in T'_p}\frac{\lost(r,t-\delay
    pr)}{\into(r,t-\delay pr)}} \\
    & \ge & \sum_p \rcvd^*(p) \,c\,|T_p|\val(p) \\
    &  = & c\,\opt.
  \end{eqnarray*}
\end{proof}

\begin{lemma}\label{lemma:aggressive}
  Let $U_p$ be the most received in any round on path $p$.
  The Linear MIMD Protocol satisfies, for each path $p$,
  \begin{eqnarray*}
  \sum_{t\in T'_p}\lsr(p,t)
  & \ge & \frac{\alpha_p(1-\beta_p)}{\beta_p(1+\alpha_p)}(|T_p|-1-\delay p{}) \\
  & & \mbox{} - \frac{1-\beta_p}{\beta_p}
  (1+\delay p{})\ln\frac{\beta_p U_p}{(\beta_p-\alpha_p) f_0(p)}.
  \end{eqnarray*}
\end{lemma}
\begin{proof}
  We use that
  (for $0 \le a \le \alpha$
  and $0 \le b\le\beta<1$),
  \[1+a-b \,\ge\, (1+a)(1-b) \,\ge\, \exp[a/(1+\alpha)  - b/(1-\beta)].\]
  
  Fix a path $p$.  Recall $\lsr(p,t) =
  \lost(p,t)/\sent(p,t-\delay p{})$.  By the definition of the
  protocol and the stated inequality, for $t = s_p+1+\delay
  p{},\ldots, e_p$,
  \begin{eqnarray*}
  \frac{\sent(p,t)}{\sent(p,t-1-\delay p{})}
  & = &  1+\alpha_p-\beta_p \lsr(p,t-1)\\
  & \ge &  
  \exp\bigg[\frac{\alpha_p}{1+\alpha_p}-\frac{\beta_p\lsr(p,t-1)}{1-\beta_p}\bigg].
  \end{eqnarray*}
  Taking logs and summing over $t$,
  \begin{eqnarray*}
  \lefteqn{\sum_{t=s_p+1+\delay p {}}^{e_p}%
           \ln\frac{\sent(p,t)}{\sent(p,t-1-\delay p{})}} \\
  & \ge & \sum_{t=s_p+1+\delay p{}}^{e_p}
  \frac{\alpha_p}{1+\alpha_p}-\lsr(p,t)\frac{\beta_p}{1-\beta_p}.
  \end{eqnarray*}
  The sum on the left-hand side telescopes.  Since at most
  $U_p$ is received on $p$ in any round, a proof by induction
  shows that at most $U_p\beta_p/(\beta_p-\alpha_p)$ is ever
  sent on $p$ in any round.  Thus,
  \begin{eqnarray*}
  \lefteqn{(1+\delay p{})\ln\frac{U_p\beta_p}{(\beta_p-\alpha_p)f_0(p)}} \\
  & \ge & 
  (|T_p|-\delay p{} - 1)  \frac{\alpha_p}{1+\alpha_p}
  -\sum_{t=s_p+1+\delay p{}}^{e_p}\lsr(p,t)\frac{\beta_p}{1-\beta_p}.
  \end{eqnarray*}

  The lemma follows.
\end{proof}

Next we combine the three lemmas.
\begin{lemma}\label{lemma:wrap}
  Fix $\epsilon>0$.
  Assume $\epsilon$-fair loss.
  Let $U_p$ be the maximum amount received in any round on path $p$.
  Define $b=\min_p (\beta_p/\alpha_p-1)\val(p)$.  Define
  \begin{eqnarray*}
  c 
  & = & \min_p
  \frac{\alpha_p(1-\beta_p)}{\beta_p(1+\alpha_p)\val(p)}(1-\frac{1+\delay p{}}{|T_p|})
  \\
  & & \mbox{} - \frac{1-\beta_p}{\beta_p\val(p)}
  \frac{1+\delay p{}}{|T_p|}\ln\frac{\beta_p U_p}{(\beta_p-\alpha_p) f_0(p)}.
  \end{eqnarray*}
  Then the total weighted throughput achieved by the protocol
  by round $T$ is at least
  \[
  b\,c\,\opt/(1+\epsilon) -\sum_p \frac{(1+\delay p{})f_0(p)\val(p)}{\alpha_p}
    \]
\end{lemma}
\begin{proof}
  Applying Lemma~\ref{lemma:conservative} and the choice of $b$,
  the total weighted throughput is
  \begin{eqnarray}    
    \lefteqn{\sum_p\sum_{t\in T'_p} \val(p)\rcvd(p,t)} \nonumber \\
    & \ge  & b\sum_p 
    \bigg(\sum_{t\in T'_p}\lost(p,t)- 
    \frac{(1+\delay p{})f_0(p)\val(p)}{\alpha_p} \bigg).
  \label{eq:1}
  \end{eqnarray}
Applying  Lemma~\ref{lemma:aggressive} and the choice of $c$,
  \begin{equation}
    \label{eq:2}
    \sum_{t\in T'_p}\lsr(p,t) \ge c |T_p|\val(p).    
  \end{equation}
$\epsilon$-fair loss over period $T'_p$ means that, for each $p$,
  \begin{equation}
    \label{eq:3}
    \sum_{t\in T'_p}\lsr(p,t) \,\le\,(1+\epsilon)
    \sum_{t\in T'_p}\sum_{r\sim p}
    \frac{\lost(r,t-\delay pr)}{\into(r,t-\delay pr)}.
  \end{equation}
Combining~(\ref{eq:2}) and~(\ref{eq:3}),
  \begin{displaymath}
    \sum_{t\in T'_p}\sum_{r\sim p}\frac{\lost(r,t-\delay p{})}{\into(r,t-\delay p{})}
    \,\ge\,
    c|T_p|\val(p)/(1+\epsilon).
  \end{displaymath}
This means the condition of Lemma~\ref{lemma:dual} is met, so
  \begin{displaymath}
    \sum_{t}\sum_{r}\lost(r,t) \ge c\,\opt/(1+\epsilon)
  \end{displaymath}
Since every lost packet is eventually observed lost on a path,
  \begin{displaymath}
    \sum_{p}\sum_{t\in T'_p} \lost(p,t) \ge \sum_{t}\sum_{r}\lost(r,t).
  \end{displaymath}
Substituting the previous two inequalities into~(\ref{eq:1}),
  the total weighted throughput is at least
  \begin{displaymath}    
    b\,c\,\opt/(1+\epsilon) \,-\, \sum_p \alpha_p^{-1}(\delay p{} + 1)f_0(p)\val(p).
  \end{displaymath}
\end{proof}

\begin{theorem}\label{theorem:main}
\mainresult
\end{theorem}
\begin{proof}
  Apply Lemma~\ref{lemma:wrap}.
  With these choices for $\alpha_p$ and $\beta_p$,
  in that lemma,
  $b=\epsilon(1-O(\epsilon))$,
  $c=\epsilon^{-1}(1-O(\epsilon))$,
  and $\sum_p(1+\delay p{})f_0(p)\val(p)/\alpha_p = O(\epsilon \opt)$.
  Thus, the lower bound that lemma gives on the weighted
  throughput is $[1-O(\epsilon)]\opt$.
\end{proof}

\Section{Discussion}
\paragraph*{Limitations of the model.}

This paper assumes static packet latencies (round-trip times).
In practice, a U.S. coast-to-coast packet round-trip time is
on the order of tens of milliseconds.  A typical router queue
might hold about 1/10 seconds worth of packets.  So, the
variation in round-trip time due to queuing can be larger than
the round-trip time itself --- our assumption of static
latencies is not realistic.  How will this affect the protocol
in practice?  Can one generalize the analysis to show that 
some amount of queuing doesn't hurt, or can queuing
destabilize the protocol?  (For one discussion of the effects
of large queues in the Internet, see \cite{morris99scalable}.)
One workaround is to take the time per iteration in the
protocol to be on the order of 1/10 second to 1 second ---
long enough so that, even with queuing, round-trip times are
likely to be only one iteration.

Here we've assumed all packets are the same size.  This hides
two practical issues: typically routing has a per-packet cost,
and large packets tend to be more likely to be dropped by
congested routers (as routers queue packets in different
queues by size).

In the current Internet, although long-lasting connections
carry most of the traffic, most connections are short-lived.
Do short-lived connections interfere?

The ``fluid packet model'' -- in which we assume that
arbitrary sending rates are possible, is unrealistic in the
case that the sending rate on a path becomes very low.  To
have confidence that the analysis in this paper is realistic,
each path should be sending at least one packet per iteration.
(In practice, it may be useful to modify the protocol so that
each path sends at least this minimal rate.)  Can one modify
the analysis to model discrete packets?

We do consider the ``fair loss'' assumption here realistic
(modulo the problem of low sending rates).  We note that
existing theoretical analyses of end-to-end protocols
typically assume fair loss at each point in time (that is,
that~(\ref{eq:fairLossExact}) holds exactly or approximately
in each round).  Other models are plausible --- for example,
random discards \cite{floyd93random} as in random early
detection (RED) --- and would allow similar assumptions.

We proved performance guarantees with respect to the {\em
  static} optimum (i.e., the sending rate on a path during its
active interval is constant).  How much does this restrict
$\opt$ in a network where capacities change slowly?

\paragraph*{Interesting directions.}

Prove upper and lower bounds on best-possible convergence
rates of on-line, end-to-end protocols that approximately
optimize various objective functions.

For example, it is straightforward to prove that any protocol
requires $\Omega(\log[\log(U/f_0)/\epsilon])$ rounds to
estimate the capacity of a single link of capacity $U$ within
a factor of $1+\epsilon$, and $1/\epsilon$ times this many
rounds to achieve an average throughput of $1-\epsilon$ times
$U$.  If we restrict the protocol so that it can't send more
than $1+O(\epsilon)$ times what it has successfully
transmitted, then these lower bounds can be improved to
$\Omega(\log(U/f_0)/\epsilon)$ rounds to estimate the capacity
of the link and $1/\epsilon$ times that to achieve an average
throughput of $1-\epsilon$ times $U$.  Strengthen these lower
bounds.

Lower bounds on related algorithms have been proven in the
contexts of static optimization \cite{klein99number} and
learning theory \cite{freund99adaptive}.  For some lower
bounds for particular protocols (e.g.\ a comparison of MIMD
and AIMD convergence on a single-path network), see
\cite{gorinsky00additive,chiu89analysis}.

An important objective function to study is {\em proportional
  fairness} (e.g.\ $\sum_p\log{}f(p)$).  For upper bounds, a
natural starting point would be to first develop a
Lagrangian-relaxation algorithm that solves the underlying
optimization problem with a provably good running time
\cite{jansen02approximation,grigoriadis94fast}.

Find a protocol that simultaneously maximizes aggregate
throughput within a constant factor {\em and} maximizes
proportional fairness within an additive constant.

A natural objection to the model in this paper is that, in
practice, most connections are open not for a given period of
time, but rather until a given number of bits have been
transmitted.  What is a reasonable objective function to model
this?  Perhaps an objective functions related to scheduling
--- e.g.\ minimize the total wait (sum of transfer times).

\section*{Appendix}
\paragraph*{Connection to existing algorithms.}

We briefly sketch the connection between the protocols
described here and existing packing and covering algorithms
\cite{garg98faster,young01sequential}.  Understanding this
connection may be useful to aid further research.  We designed
the protocol in this paper by starting with
Lagrangian-relaxation algorithms for the dual problem (a
covering problem --- fractional weighted set cover) of the
underlying static multicommodity flow problem, and then
adapting it to the current setting (e.g.\ with the fair loss
assumption, round-trip times, etc.).  The resulting algorithm
is by no means a direct translation of an existing algorithm
--- in fact, appropriately parameterized, it runs faster than
existing algorithms as they are implemented in the literature.
However, some connections remain.  Briefly, each round of the
protocol corresponds to an iteration of the algorithm.  Each
path $p$'s sending rate is proportional to a path variable
$x_p$ in the algorithm.  Each resource $r$'s cumulative packet
loss is proportional to a resource variable $\ell_r$ in the
algorithm.  Roughly, the algorithm ensures that (by time $T$)
we get something like the familiar invariant \(x_p \approx
(1-\epsilon)^{\sum_{r\sim p} \ell_r}\).  When a resource
discards a packet, the resulting adjustments in the sending
rates correspond to incrementing the corresponding path
variable; that only congested resources lose packets means
that resource variables get incremented only if their
corresponding constraints are ``sufficiently violated''.

\bibliographystyle{latex8}
\bibliography{new,tcp,lagrange,bandwidth}

\end{document}